\documentstyle[11pt]{article}
\def\be{\begin{equation}}
\def\ee{\end{equation}}
\def\bea{\begin{eqnarray}}
\def\eea{\end{eqnarray}}
\def\e{\rm {e}}
\def\ep{\epsilon}
\def\l{\cal L}
\begin{document}
\begin{flushright} hepth@xxx/9704037 \\  CERN-TH/97-63\\  
Miami TH/1/97 \\  ANL-HEP-PR-97-10  \end{flushright}
{\Large
\centerline{{\bf Integrable Symplectic Trilinear Interaction Terms for 
Matrix Membranes}}

Thomas Curtright$^{\S}$, David Fairlie$^\natural$\footnote{ On research leave 
from the University of Durham, U.K.}, and Cosmas Zachos$^{\P}$\\ }

$^{\S}$ Department of Physics, University of Miami,
Box 248046, Coral Gables, Florida 33124, USA\\
\phantom{.} \qquad\qquad{\sl curtright@phyvax.ir.Miami.edu}  

$^{\natural}$ Theory Division, CERN, CH-1211 Geneva 23, SWITZERLAND\\ 
\phantom{.} \qquad\qquad{\sl David.Fairlie@cern.ch} 

$^{\P}$ High Energy Physics Division,
Argonne National Laboratory, Argonne, IL 60439-4815, USA \\
\phantom{.} \qquad\qquad{\sl zachos@hep.anl.gov}      

\begin{abstract}
Cubic interactions are considered in 3 and 7 space dimensions, respectively,
for bosonic membranes in Poisson Bracket form. Their symmetries and vacuum
configurations are discussed. Their associated first order equations are
transformed to Nahm's equations, and are hence seen to be integrable, for the
3-dimensional case, by virtue of the explicit Lax pair provided. The
constructions introduced also apply to commutator or Moyal Bracket analogues. 
\end {abstract}
\noindent \rule{7in}{0.1em}
\vskip 0.4cm

\noindent {\bf Introduction}

A proposal for non-perturbative formulation of 
M-theory \cite{banks} has led to a revival of matrix membrane theory
\cite{collins,hoppe1}. Symmetry features of membranes and their 
connection to matrix models \cite{hoppe1,floratos3,ffz,floratos2,ffzjmp} have 
been established for quite some time. Effectively, infinite-$N$ quantum 
mechanics matrix models presented as a restriction out of SU$(\infty)$ 
Yang-Mills theories 
amount to membranes, by virtue of the connection between SU($N$)
and area-preserving diffeomorphisms ({\em Sdiff}) generated by Poisson 
Brackets, in which ``colour" algebra indices Fourier-transform to ``membrane" 
sheet coordinates. The two are underlain and linked by Moyal Brackets.

Below, we consider novel Poisson Bracket interactions for a bosonic 
membrane embedded in 3-space
 \be
 {\l}_{IPB} = \frac{1}{3} \ep^{\mu\nu\kappa} X^\mu \{ X^\nu ,\ X^\kappa \},
 \label{termPB}
 \ee
which are restrictions of the Moyal Bracket generalization
 \be
 {\l}_{IMB} = \frac{1}{3} \ep^{\mu\nu\kappa} X^\mu \{ \{ X^\nu ,\ X^\kappa \}\},
 \label{termM}
 \ee
which, in turn, also encompasses the plain matrix commutator term
 \be
 {\l}_{IC} = \frac{1}{3} \ep^{\mu\nu\kappa} X^\mu [ X^\nu ,\ X^\kappa ]~.
 \label{termC}
 \ee

The structure of (\ref{termPB}) may be recognized as that of the interaction 
term   $\ep_{ijl} \phi^i \ep^{\mu\nu}  \partial_\mu \phi^j  
\partial_\nu \phi^j $ 
of the 2-dimensional SO(3) pseudodual chiral $\sigma$-model of Zakharov and 
Mikhailov \cite{zakharov}---this is a limit \cite{pascos} of the WZWN 
interaction term, where the integer WZWN coefficient goes to infinity while 
the coupling goes to zero, such that the product of the integer with the cube 
of the coupling is kept constant. 
The structure of (\ref{termC}) is also linked to what remains of the gauge 
theory instanton density,
\be 
K^0 = \ep^{\mu\nu\kappa} {\rm Tr} A_\mu \left( \partial_{\nu}A_{\kappa} - 
\frac{1}{4} [A_{\nu},A_{\kappa}]\right)~, 
\ee
in the standard space-invariant limit (where the first term vanishes). (N.B.\ 
This interaction may be contrasted to the one which appears in a different
model \cite{plebanski}.) There is some formal resemblance to membrane
interaction terms introduced in \cite{zaikov}, in that case a quartic in the
$X^\mu$'s, which, in turn, reflect the symplectic twist of topological terms
\cite{biran} for self-dual membranes. Unlike those interactions, the cubic
terms considered here do not posit full Lorentz invariance beyond 3-rotational
invariance: they are merely being considered as quantum mechanical systems with
internal symmetry. One may expect this fact to complicate supersymmetrization. 

We then also introduce analogous trilinear  interactions for membranes embedded
in 7-space, which also evince similarly interesting properties. 

In what follows, after a brief review of some matrix membrane technology, we
discuss the symmetry features of the new terms, the symmetry of the
corresponding vacuum configurations, and describe classical configurations of
the Nahm type, which we find to be integrable, as in the conventional membrane
models. Our discussion will concentrate on Poisson Brackets, but the majority
of our results carry over to the Moyal Bracket and commutator cases, by dint of
the underlying formal analogy. Subtler considerations of special features for
various membrane topologies are not discussed here, and may be addressed as in
\cite{kim}. Nontrivial boundary terms, e.g.~of the type linked to D-branes, are
not yet addressed. 
 
\noindent {\bf Review and Notation } 

Poisson Brackets, Moyal Brackets, and commutators are inter-related derivative
operators, sharing similar properties such as Leibniz chain rules, integration
by parts, associativity (so they obey the Jacobi identity), etc. Much of their
technology is reviewed in \cite{moyal,ffz,ffzjmp,hoppe2}. 

Poisson Brackets act on the ``classical phase-space" of  Fourier-transformed 
colour variables, with membrane coordinates $\xi=\alpha,\ \beta$, 
\be
\{X^\mu,\ X^\nu\}=\frac{\partial X^\mu}{\partial\alpha}\frac{\partial 
X^\nu}{\partial\beta}-
 \frac{\partial X^\mu}{\partial\beta}\frac{\partial X^\nu}{\partial\alpha} ~.
 \label{PB}
\ee
This might be effectively regarded as the infinitesimal 
canonical transformation on the coordinates $\xi$ of $X^\nu$, 
generated by $\nabla X^\mu\times \nabla $, 
s.t.\ $(\alpha,\beta) \mapsto (\alpha- \partial X_\mu/\partial \beta~,~\beta
+\partial X_\mu /\partial \alpha)$,
which preserves the membrane area element $d\alpha d\beta $. This element 
is referred to as a {\it symplectic form} and the class of  transformations 
that leaves it invariant specifies a symplectic geometry; the area preserving
diffeomorphisms are dubbed {\em Sdiff} . 

PBs correspond to $N\rightarrow\infty$ matrix commutators.
But there is a generalization which covers both finite and 
infinite $N$. The essentially  unique associative generalization of PBs is the 
Moyal Bracket \cite{moyal}, 
 \be
 \{ \{ X^\mu,\ X^\nu \} \} =   \frac{1}{\lambda}  \sin\left( \lambda 
\frac{\partial }{\partial\alpha}\frac{\partial }{\partial\beta'}-\lambda 
 \frac{\partial }{\partial\beta}\frac{\partial }{\partial\alpha'} \right) ~
X^\mu (\xi)   X^\nu (\xi')  \Biggl|_{\xi'=\xi}    ~.
\label{moyal}
 \ee
For $\lambda= 2\pi /N $, ref \cite{ffz} demonstrates that the Moyal bracket is 
essentially equivalent to the commutator of SU($N$) matrices (or subalgebras of
SU($N$), depending on the topology of the corresponding membrane surface
involved in the Fourier-transform of the colour indices \cite{ffzjmp,kim}). In
the limit $\lambda \rightarrow 0$, the Moyal bracket goes to the PB (i.e.\ 
$\lambda$ may be thought of as $\hbar$). Thus, PBs are seen to represent the
infinite $N$ limit. This type of identification was first noted by Hoppe on a
spherical membrane surface \cite{hoppe1}; the foregoing Moyal limit argument
was first formulated on the torus \cite{ffz}, but extends naturally to other
topologies \cite{ffz,ffzjmp,kim}. 

Refs \cite{floratos3} utilize the abovementioned identification of 
SU($\infty$) with {\em Sdiff}  on a 2-sphere, to take the large $N$ limit of 
SU($N$) gauge theory and produce membranes. This procedure was found to be
more transparent on the torus \cite{ffzjmp}: the Lie algebra indices
Fourier-conjugate to surface coordinates, and the fields are rescaled Fourier
transforms of the original SU($N$) fields. The group composition rule for them
is given by the PBs and the group trace by surface integration, 
\be
[A_{\mu}, A_{\nu} ] ~\mapsto  \{ a_{\mu},a_{\nu} \} ~;
\ee
\be
F_{\mu \nu}=\partial_{\mu}A_{\nu}-\partial_{\nu}A_{\mu} +[A_{\mu}, A_{\nu}
]~\mapsto 
f_{\mu\nu}(\alpha, \beta) =\partial_{\mu}a_{\nu}-\partial_{\nu}a_{\mu}
+\{ a_{\mu},a_{\nu}\} ~;
\ee
\be
\hbox{Tr} F_{\mu\nu}F_{\mu\nu}\mapsto -{N^3\over 64\pi^4} \int \! d\alpha 
d\beta  ~f_{\mu\nu}(\alpha ,\beta ) ~f_{\mu\nu} (\alpha ,\beta ).
\ee

But the large $N$ limit need not be taken to produce sheet actions. 
The Lagrangian with the Moyal Bracket supplanting the Poisson Bracket is
itself a gauge-invariant theory, provided that the gauge transformation also 
involves the Moyal instead of the Poisson bracket:
\be
\delta a_{\mu} =\partial_{\mu} \Lambda-\{\{\Lambda,a_{\mu}\}\}  ~,
\ee
and hence, by virtue of the Jacobi identity,
\be
\delta f_{\mu\nu}=-\{\{\Lambda,f_{\mu\nu}\}\}      ~.
\ee
colour invariance then follows, 
\be
\delta\int \! d\alpha d\beta ~f_{\mu\nu} f_{\mu\nu}=
-2\int \! d\alpha d\beta ~f_{\mu\nu}\{\{\Lambda,f_{\mu\nu}\}\}=0~.
\ee
The relevant manipulations are specified in \cite{ffzjmp}, and the  
last equality is evident by integrations by parts, where the surface term
is discarded, or nonexistent if the colour membrane surface is closed.
(But note this topological term may be nontrivial for D-membranes.)
For $\lambda= 2\pi /N $, this is equivalent to the conventional SU($N$) 
commutator gauge theory.

Consider the SU$(\infty)$ Yang Mills Lagrangian and dimensionally reduce 
all space dependence \cite{hoppe1}, leaving only time dependence, while 
preserving all the colour-Fourier-space 
(membrane coordinates $\xi=\alpha,~\beta$) dependence of the gauge fields, 
which are now denoted $X^\mu(t,\alpha,\beta)$. Fix the gauge to $X^0=0$.
The Yang-Mills lagrangian density reduces to the bosonic membrane 
lagrangian density
\be
 {\l}_{PB} = \frac{1}{2}(\partial_t X^\mu)^2 -\frac{1}{4}\{X^\mu,\ X^\nu\}^2 .
 \label{legba}
\ee
 The PB is also 
the determinant of the tangents to the membrane, so the conventional 
``potential term" was identified in \cite{ffzjmp} as the Schild-Eguchi string 
lagrangian density \cite{schild} (sheet area squared instead of area), 
$ \{ X_{\mu}, X_{\nu}\} \{ X_{\mu}, X_{\nu}\}$. (It can be seen that the 
equations of motion of such a string action contain those of Nambu's action.) 

Note that, fixing the gauge $X_0=0$ preserves the global colour invariance,
i.e.~with a time-independent parameter $\Lambda(\alpha,\beta)$.  The action is 
then  invariant under 
\be
 \delta X^\mu = \{\Lambda ,\ X^\mu\} .  \label{hatchibombotar}
\ee
By Noether's theorem, this implies the time invariance of the colour charge, 
\be
 {\cal Q}_\Lambda   = \int\! d\alpha d\beta ~\Lambda(\alpha,\beta) ~
\{\partial_t X^\mu ,\ X^\mu\} .
\ee

The same also works for the Moyal case \cite{ffzjmp}. The corresponding Moyal 
Schild-Eguchi term was utilized to yield a ``star-product-membrane'' 
\cite{hoppe2},
 \be
 {\l}_{MB}=\frac{1}{2}(\partial_t X^\mu)^2 -\frac{1}{4}\{\{X^\mu,\ X^\nu\}\}^2,
 \label{bondye}
\ee
invariant under 
\be
 \delta X^\mu = \{ \{ \Lambda ,\ X^\mu \} \} .
 \label{odoun}
\ee
As outlined, this includes the commutator case, 
\be
 {\l}_{C} = \frac{1}{2}(\partial_t X^\mu)^2 -\frac{1}{4} [X^\mu,\ X^\nu]^2,
 \label{doulamor}
\ee
invariant under 
\be
 \delta X^\mu = [\Lambda ,\ X^\mu ].
 \label{varC}
\ee

\noindent {\bf Discussion of the Cubic Terms for 3 Dimensions}

By suitable integrations by parts, it is straightforward to check that the 
cubic terms (\ref{termPB},\ref{termM},\ref{termC}) in the respective actions, 
$\int dt d\alpha d\beta ~\l$, are 3-rotational invariant, as well as
time-translation invariant and translation symmetric. They are also 
global colour invariant, as specified above. 

Let us also consider a plain mass term in the action, (of the type that may 
arise as a remnant of space gradients in compactified dimensions), 
\be
 {\l}_{3dPB} = \frac{1}{2}(\partial_t X^\mu)^2 -\frac{1}{4}\{X^\mu,\ X^\nu\}^2
 -\frac{m}{2} \ep^{\mu\nu\kappa} X^\mu \{ X^\nu ,\ X^\kappa \} -
 \frac{m^2}{2}(X^\mu)^2  .
 \label{bokor}
\ee
The second order equation of motion,
\be
 \partial_t^2 X^\mu  = - m^2 X^\mu -\frac{3m}{2}\ep^{\mu\nu\kappa} 
 \{ X^\nu ,\ X^\kappa \}-  \{  X^\nu  ,  \{ X^\mu ,\ X^\nu \}\} ,
 \label{bossou}
\ee
follows not only from extremizing the action, but also results from a 
first-order equation of the Nahm (self-dual) type \cite{neq}, 
albeit complex, 
\be
 \partial_t X^\mu = imX^\mu +\frac{i}{2}\ep^{\mu\nu\kappa} \{ X^\nu ,\ 
 X^\kappa \}.
 \label{nahm}
\ee
These equations hold for PBs, MBs, and commutators. 

For solutions of this first-order  equation, the conserved energy vanishes.
In general, however, such solutions are not real, and do not provide 
absolute minima for the action. Nonetheless, the lagrangian density can be 
expressed as a sum of evocative squares, since the
potential in (\ref{bokor}) is such a sum, 
\be
 {\l}_{3dPB} = \frac{1}{2}(\partial_t X^\mu)^2 -\frac{1}{2}\left( m X^\mu 
 +\frac{1}{2}\ep^{\mu\nu\kappa} \{ X^\nu ,\ X^\kappa \}\right) ^2.\label{mambo} 
\ee
By integration by parts in the action $\int dt d\alpha d\beta ~ {\l}_{3dPB} $,
the lagrangian density itself can then be altered to 
\be
 {\l}_{3dPB}\cong - \frac{1}{2}\left( i\partial_t X^\mu +m X^\mu +\frac{1}{2}
 \ep^{\mu\nu\kappa} \{ X^\nu ,\ X^\kappa \}\right) ^2 ,
\ee 
just like the conventional bosonic membrane lagrangian density (the congruence 
symbol, $\cong$, indicates equivalence up to surface terms, which, e.g., 
vanish for a closed surface; again, consideration of D-membranes would 
proceed separately). The complex-conjugate versions of the above are 
equally valid.       

\noindent{\bf Vacuum Configurations} 

The minimum of the conventional matrix 
membrane trough potential favors alignment of the $X^\mu$s. The mass 
parameter introduced above parameterizes a partial trough symmetry 
breaking \cite{curtright}, but does not lift ``dilation" invariance, seen as 
follows. 

The static ($t$-independence) minima for the action 
(vacuum configurations) are solutions of
\begin{equation}
 mX^{\mu }+\frac{1}{2}\epsilon ^{\mu \nu \kappa }\{X^{\nu },\ X^{\kappa}\}
 =0\;.  \label{static}
\end{equation}
The previously considered case, $m=0$, is easily solved by ``colour-parallel"
configurations. But for $m\neq 0$, the static solutions must lie on a 
2-sphere, since from the previous equation
\begin{equation}
 X^{\mu }\frac{\partial X^{\mu }}{\partial \alpha }=0=X^{\mu }\frac{\partial
 X^{\mu }}{\partial \beta }\;,
\end{equation}
so 
\be
X^{\mu }X^{\mu }=R^2~,
\ee
an unspecified  constant. (Hence 
$\epsilon^{\mu \nu \kappa } X^\mu \{ X^{\nu },\ X^{\kappa} \}=-2mR^2$.)
However, from (\ref{static}), note that both $m$ and 
also $R$, the scale of the $X^\mu$s, can
be absorbed in the membrane coordinates $\xi$ and will not be specified by
the solution of (\ref{static}).  

Indeed, solving for one 
coordinate component on this sphere, say 
\be
X(Y,Z)=\pm  \sqrt{R^{2}-Y^{2}-Z^{2}}~,
\ee
reduces the three equations (\ref{static}) to one. Namely, 
\begin{equation}
 \{Z,Y\}=m\sqrt{R^{2}-Y^{2}-Z^{2}}~,
\end{equation}
on the positive $X$ branch ($m\mapsto -m$ on the negative $X$ branch). This
last equation is solved by 
\be
 Z=\alpha , \qquad Y=\sqrt{R^{2}-\alpha ^{2}}\sin (m\beta )~.
\ee
One can then interpret $m\beta $ as the usual azimuthal angle
around the $Z$-axis. Hence, $-\pi /2\leq m\beta \leq \pi /2$ and $-R\leq
\alpha \leq R$ covers the $X\geq 0$ hemisphere completely. The other
hemisphere is covered completely by the negative $X$ branch. Since $R$ is
not fixed, it amounts to an unlifted residual trough dilation degeneracy.

All static solutions are connected to this explicit one by
rescaling $R$ and exploiting the equation's area-preserving diffeomorphism
invariance for $\xi =(\alpha ,\beta )$.

\noindent{\bf Nahm's Equation and Lax Pairs} 

The first order equation, (\ref{nahm}), 
simplifies upon changing variables to $\tau =\e^{imt}/m  $ and
$X^\mu =\e^{imt} Y^\mu$, and reduces to the conventional PB
version \cite{ward} of Nahm's equation \cite{neq},  
\be
 \partial_\tau Y^\mu  =  \frac{1}{2}\ep^{\mu\nu\kappa} \{ Y^\nu ,\ Y^\kappa \}.
 \label{orignahm}
\ee
This has real solutions, and can be linearized by Ward's transformation 
\cite{ward}. However, the action (\ref{mambo}) does not reduce to the 
conventional one upon these 
transformations. (Likewise, the second order equations of motion
only reduce to 
$ \partial_{\tau}^2  Y^\mu  =  \{ Y^\nu, \{ Y^\mu , Y^\nu \} \}+  
\frac{3}{\tau}( \partial_\tau Y^\mu  - \frac{\ep^{\mu\nu\kappa}} {2}  
\{ Y^\nu , Y^\kappa \})$.)  

Moreover, note 
\be
\partial_\tau Y^\mu \partial_\tau Y^\mu = { \partial (Y^1 ,Y^2, Y^3)\over 
\partial (\tau ,\alpha,\beta)} ,
\ee
\be
\partial_\tau Y^\mu \partial_\xi Y^\mu =0 .
\ee
 
One may further utilize the cube root of unity, $\omega =\exp (2\pi i/3)$,
(note $\omega (\omega -1)$ is pure imaginary) to recast (\ref{orignahm}),
\be
 L\equiv \omega  Y^1 + \omega ^2 Y^2 + Y^3~, \qquad 
 \overline{L}=\omega^2  Y^1 + \omega Y^2 + Y^3~, \qquad 
 M\equiv Y^1 + Y^2 + Y^3~, 
\ee
\be
 \omega (\omega -1)~ \partial_\tau L= \{  M~,~ L\}~,\qquad 
 \omega (\omega -1)~ \partial_\tau \overline{L}= 
 -\{  M~,~ \overline{L} \}~, \qquad 
 \omega (\omega -1)~ \partial_\tau M= \{ L ~,  \overline{L} \}~,  \label{erzulie}
\ee
which thus yields an infinite number of complex time-invariants,
\be 
Q_n=\int\! d\alpha d\beta ~L^n   ~,  
\ee
for arbitrary integer power $n$, as the time derivative of the integrand 
is a surface term. (This is in complete analogy with the standard case of 
commutators.)  This is linked to classical integrability, as discussed next.

Eqs (\ref{erzulie}) amount to one complex and one real equation, but 
these can further be compacted into just one by virtue of an arbitrary real 
spectral parameter $\zeta$, introduced in \cite{leontaris}: 
\be
 H\equiv {i\over \sqrt{2} \omega(\omega-1)}
 \left( \zeta L - {\overline{L} \over \zeta}\right) ~, \qquad \qquad 
 K\equiv i \sqrt{2} M + \zeta L + {\overline{L} \over \zeta} ~, 
\ee
\be
 \partial_\tau K= \{ H, K \}~.
\ee
(Note the wave solution $H=\alpha,~ L= f(\beta + \tau)$.)
This Lax pair, analogous to \cite{leontaris},  
likewise leads to a one-parameter family of time-invariants,
\be
{\cal Q}_n (\zeta) =\int\! d\alpha d\beta ~ K^n ~,  
\ee
and a Lax isospectral flow, 
\be
{\cal K}\equiv \nabla K \times \nabla ,\quad \qquad 
{\cal H}\equiv \nabla H \times \nabla ,\quad \qquad 
 \nabla \equiv \left( {\partial \over \partial\alpha}, 
{\partial\over \partial\beta}\right) ~, 
\ee
such that:
\be 
\partial_\tau  {\cal K}  = {\cal H} {\cal K} -{\cal K} {\cal H} ~.
\ee
As a consequence, the spectrum of ${\cal K}$ is preserved upon time evolution 
by the (pure imaginary) ${\cal H}$:
\be
\partial_\tau \psi= {\cal H} \psi,
\ee
since time-differentiating 
\be
{\cal K} \psi= \lambda\psi 
\ee
and applying the above yields 
\be  
(\partial_\tau \lambda) ~\psi = (\partial_\tau  {\cal K}) \psi+ 
{\cal K}\partial_\tau  \psi- \lambda~\partial_\tau \psi =0  ~.
\ee 
This isospectral flow then provides integrability \cite{zakharov} 
for (\ref{nahm}), as in the case of the conventional Nahm equation.

This discussion also carries over to plain commutators or Moyal Brackets 
as well, with suitable adaptations \cite{ffz}. 

\noindent{\bf Duality and Implicit Solution of the (PB) Nahm Equation}

Ward \cite{ward} has solved eq (\ref{orignahm}) implicitly through twistor 
 linearization.

Another solution procedure for the construction of a wide class of solutions to
(\ref{orignahm}) may be found by interchanging the r\^oles of dependent and
independent variables; the equations then take the form
\be
\frac{\partial \tau}{\partial Y^1}=\frac{\partial \alpha}{\partial Y^2}
\frac{\partial \beta}{\partial Y^3}-
\frac{\partial \alpha}{\partial Y^3}\frac{\partial \beta}{\partial Y^2} ~,
\label{replace}
\ee
together with cyclic permutations, i.e. 
\be
\partial_\mu \tau =\ep^{\mu\nu\rho} \partial_\nu \alpha ~\partial_\rho \beta ~.
\label{philophage}
\ee

Cross-differentiation produces integrability conditions 
\be
\partial_\mu ( \partial_\kappa \alpha ~\partial_\mu \beta - 
\partial_\mu \alpha ~\partial_\kappa \beta)    =0. 
\label{podular}
\ee
Another evident consistency condition is  
\be
\partial^2 \tau = 0~,
\label{houngan}
\ee
but this will be guaranteed by the integrability of those equations. 
Having found 
$\alpha,\ \beta$ in terms of $Y^1,\ Y^2,\ Y^3$, 
the solution for $\tau$ follows by quadratures. 

Evidently, solutions of these dual equations 
\be
\partial_\mu f =\ep^{\mu\nu\rho} \partial_\nu  g  ~ \partial_\rho h ~,
\label{philopote}
\ee
for $f(Y^1,Y^2,Y^3)$ produce constants of the motion $\int d^2 \xi~ f$,
beyond those already found by the Lax procedure, for 
the original equation (\ref{orignahm}), in illustration of a phenomenon 
noted in \cite{strach}, as it is straightforward to verify that 
$df/d\tau=\{g,h\}$. Actually, $f$ need only solve Laplace's equation 
(\ref{houngan}): any harmonic function $f(Y^1,Y^2,Y^3)$ yields a conserved 
density for (\ref{orignahm}), by also satisfying (\ref{philopote}). 
By virtue of Helmholtz's theorem, a divergenceless 3-vector $\nabla f$ is 
representable as a curl of another vector ${\bf A}$. On the other hand, an 
arbitrary 3-vector can also be represented in terms of
three scalars by means of the Clebsch decomposition of that vector as
${\bf A}=g\nabla h + \nabla u $.  In fact, the problem of solving the inverse 
Nahm equation (\ref{replace}) is equivalent to determining the Clebsch 
decomposition of an arbitrary vector.

A class of solutions can be found by postulating a simple dependence on
$Y^3$; with the Ansatz 
\be
\tau =f(Y^1,Y^2),\ \ \ \alpha=\e^{-mY^3}g(Y^1,Y^2),\ \ \ \beta=e^{mY^3}
h(Y^1,Y^2),
\label{postular}
\ee
it is seen that the equations are satisfied, provided $g$ is an arbitrary 
function of $h$,
\be  
g=\phi(h),
\ee
as well as 
\be
\tau=\Re F(Y^1+iY^2),\ \ \ mgh =\Im F(Y^1+iY^2)~, 
\label{cond}
\ee
for an arbitrary analytic function $F$, since these combinations must satisfy 
the Cauchy-Riemann conditions.

\noindent{\bf Membrane Embedding in 7 Dimensions}

Remarkably, the same type of term may also be introduced for a membrane 
embedded in 7 space dimensions. A self-dual (antisymmetric) 4-tensor in 8 
dimensions, $f_{\mu\nu\rho\sigma}$ was invoked \cite{cor} as an
8-dimensional analogue of the 4-dimensional fully antisymmetric tensor
$\epsilon_{\mu\nu\rho\sigma}$. Some useful technology for the manipulation
of this tensor (which has 35 nonzero components and is invariant under a 
particular $SO(7)$ subgroup of $SO(8)$) can also be found in \cite{gursey}; 
in particular, the identity
\be
 f^{0\mu\nu\kappa} f^{0\mu\lambda\rho}= f^{\nu\kappa\lambda\rho} 
 +\delta^{\nu\lambda} \delta^{\kappa\rho} -\delta^{\kappa\lambda}
  \delta^{\nu\rho}~.\label{gurseyidentity} 
\ee

By analogy with (\ref{orignahm}), one can postulate a first order equation 
\be
 \partial_\tau Y^\mu -{f^{0\mu\nu\kappa}\over 2}  \{ Y^\nu ,\ Y^\kappa \} =0~. 
 \label{orignahm2}
\ee
The indices run from $\mu =1$ to $\mu=7$, since we are working in a gauge 
where $Y^0=0$.
The second order equation arising from iteration of (\ref{orignahm2}), by 
virtue of the above identity as well as the Jacobi identity is 
\be
 \partial_{\tau\tau} Y^\mu  = -\{\{ Y^\mu ,\ Y^\nu \}  ,\ Y^\nu \}  ~.
 \label{baronsamedi}
\ee

This arises from the lagrangian density 
\be
 {\l}_{7dPB}=\frac{1}{2} (\partial_\tau Y^\mu)^2+\frac{1}{4}\{Y^\mu,\ 
  Y^\nu\}^2 ~ . \label{tapitor}
\ee
As in the 3-dimensional case, this action is a sum of squares up to a mere 
surface term,
\be
 \frac{1}{2} \left( \partial_\tau Y^\mu  -  {f^{0\mu\nu\kappa}\over 2}  
 \{ Y^\nu ,\ Y^\kappa \}\right)^2  
 ={\l}_{7dPB} - f^{0\mu\nu\kappa} \partial_\tau Y^\mu\partial_\alpha 
 Y^\nu\partial_\beta Y^\kappa \cong {\l}_{7dPB}. \label{top}
\ee
Apparent extra quartic terms in this lagrangian density have, in fact, 
vanished by virtue of the identity,
\be
 \{f,\ g\}\{h,\ k\} + \{f,\ h\}\{k,\ g\}+\{f,\ k\}\{g,\ h\}\equiv 0,
 \label{saturday}
\ee
which holds for Poisson Brackets on a 2-dimensional phase-space---but not for
commutators nor Moyal Brackets\footnote{ D. Fairlie and A. Sudbery, 1988, 
unpublished. It follows from $\epsilon^{[jk} \delta^{l]m}=0$, whence  
$\epsilon^{[jk} \epsilon^{l]m}=0$ for these membrane symplectic coordinates.}.
This cancellation works at the level of the lagrangian density for the PB case.
However, note that even for ordinary matrices the corresponding term would 
vanish in the traced action, by the cyclicity of the trace pitted against 
full antisymmetry, 
\be
f^{\mu\nu\kappa\rho }\hbox{Tr} X^\mu X^\nu X^\kappa X^\rho =0~.
\ee
Likewise, the corresponding interaction for Moyal Brackets,
\be
f^{\mu\nu\kappa\rho }\int\! d^2 \xi ~\{\{X^\mu, X^\nu\}\}
\{\{ X^\kappa, X^\rho \}\} ,
\ee
is forced by associativity to reduce to a surface term, 
vanishing unless there are contributions from surface boundaries or
D-membrane topological numbers involved. (Shortcuts for the manipulation of 
such expressions underlain by $\star$-products can be found in, 
e.g.~\cite{hoppe2}.) The cross terms involving time derivatives are 
expressible as divergences, as in the 3-dimensional case, and hence give rise 
to possible topological contributions.

As a result, (\ref{orignahm2}) is the Bogomol'nyi minimum of the action 
(\ref{tapitor}) with the bottomless potential.

As in the case of 3-space, the conventional membrane signs can now be 
considered (for energy bounded below), and a symmetry breaking 
term $m$ introduced, to yield 
\be
{\l}_{7dPB}\cong - \frac{1}{2}\left(- i\partial_t X^\mu +m X^\mu +
{f^{0\mu\nu\kappa}\over 2}  \{ X^\nu ,\ X^\kappa \}\right) ^2.
\ee 

This model likewise has 7-space rotational invariance, and its vacuum 
configurations are, correspondingly, 2-surfaces lying on the spatial 6-sphere 
embedded in 7-space: $X^{\mu }X^{\mu }=R^{2}$. But, in addition, because 
of (\ref{gurseyidentity}), these surfaces on the sphere also satisfy the
trilinear constraint 
\be
 f^{\lambda \mu \nu \kappa }X^{\mu }\{X^{\nu },\
 X^{\kappa }\}=0 ~,
\ee
for $\lambda\neq 0$. (For $\lambda=0$ this trilinear is $-2mR^2$.) 

\noindent {\bf Higher-dimensional Analogues}

The system of equations (\ref{orignahm2}) does not appear to readily 
yield integrability properties. However, it may be extended to a 9-dimensional
system---or, equivalently, a 10-dimensional one if $\partial_\tau Y^\mu$ is 
replaced by $\{Y^0 ,\ Y^\mu\}$. This different system, though it breaks the 
9-dimensional rotational invariance, is integrable by virtue of the additional 
six constraints imposed on the system.

Writing out the system (\ref{orignahm2}) explicitly and augmenting it with two
more variables, $Y^8,\ Y^9$, gives rise to the equations     \pagebreak
\bea
 \partial_\tau Y^1& = & \{Y^2 ,\ Y^3\}+\{Y^6,\ Y^5\} + \{Y^4,\ Y^7\} 
 +\{Y^8,\ Y^9\}\nonumber\\ 
 \partial_\tau Y^2& = & \{Y^3 ,\ Y^1\}+\{Y^4,\ Y^6\} +\{Y^5,\ Y^7\}\nonumber\\
 \partial_\tau Y^3& = & \{Y^1 ,\ Y^2\}+\{Y^5,\ Y^4\} +\{Y^6,\ Y^7\}\nonumber\\
 \partial_\tau Y^4& = & \{Y^3 ,\ Y^5\}+\{Y^6,\ Y^2\} +\{Y^7,\ Y^1\}\nonumber\\
 \partial_\tau Y^5& = & \{Y^4 ,\ Y^3\}+\{Y^1,\ Y^6\} +\{Y^7,\ Y^2\}\nonumber\\
 \partial_\tau Y^6& = & \{Y^2 ,\ Y^4\}+\{Y^5,\ Y^1\} +\{Y^7,\ Y^3\}\nonumber\\
 \partial_\tau Y^7& = & \{Y^1 ,\ Y^4\}+\{Y^2,\ Y^5\} +\{Y^3,\ Y^6\}\nonumber\\
\partial_\tau Y^8& = & \{Y^9,\ Y^1\}\nonumber\\
 \partial_\tau Y^9& = & \{Y^1,\ Y^8\} \nonumber\\
0&=& \{Y^8,\ Y^2\}+\{Y^3,\ Y^9\}\nonumber\\
0&=& \{Y^8,\ Y^3\}+\{Y^9,\ Y^2\}\nonumber\\
0&=& \{Y^8,\ Y^6\}+\{Y^5,\ Y^9\}\nonumber\\
0&=& \{Y^8,\ Y^5\}+\{Y^9,\ Y^6\}\nonumber\\
0&=& \{Y^8,\ Y^4\}+\{Y^7,\ Y^9\}\nonumber\\
0&=& \{Y^8,\ Y^7\}+\{Y^9,\ Y^4\}~.
\label{orignahm3}
 \eea

The sums of the squares of these expressions amount to the sums of the
squares of the individual terms, each appearing once and once only, with unit
coefficient, the cross terms vanishing by the identity (\ref{saturday}).
Other schemes are also possible \cite{ward2}, but in the 10-dimensional cases
investigated \cite{dbf},  the quadratic terms do not arise with both unit 
multiplicity and positive sign. Whilst the cases considered in the latter
reference were directly associated with the reduction of $N=1$ supersymmetric
10-dimensional Yang Mills to the $N=4$ 4-dimensional case, the symmetry
breakdown in (\ref{orignahm3}) does not follow this pattern.

The terms not involving a $\tau$-derivative in (\ref{orignahm3}) may be
rearranged as follows,
\be
 \{(Y^2\pm i Y^3),\ (Y^8\pm i Y^9)\}=\{(Y^6\pm i Y^5),\ (Y^8\pm i Y^9)\}=
 \{(Y^4\pm i Y^7),\ (Y^8\pm i Y^9)\}=0.
 \label{complex}
\ee
This means that these ``commuting" combinations of variables are functionally 
related:
\be
(Y^2+ i Y^3)=A_{+}(\alpha,\beta,\tau),~~ (Y^6 + i Y^5)=
B_{+}(A_{+}),~~ (Y^8+ i Y^9)=C_{+}(A_{+} ),~~ (Y^4+ i Y^7)=D_{+}(A_{+} ) ,
\label{function}
\ee
with a corresponding result for the negative combinations.
This implies that all $\tau$ derivative equations, apart from the first, are 
equivalent to the Lax pair combinations
\be
\partial_\tau (Y^2\pm iY^3) =\pm i\{Y^1,\ (Y^2\pm iY^3)\},
\label{loa}
\ee  
i.e.~the 6 constraints have reduced 9 equations to 3.
The arbitrariness in the functions $B_{\pm}, C_{\pm},\ D_{\pm}$, implies that 
the first equation in (\ref{orignahm3}) is less restrictive than the 
corresponding case of (\ref{orignahm}). Thus (\ref{orignahm3}) are a fortiori 
integrable, and the solution is not dissimilar to the 3-dimensional case.

Even though mostly integrable first-order equations have been studied 
in this note, it should be borne in mind that the behaviour of the generic 
solutions to the second-order equations of motion for such systems is often 
chaotic. For example, in the case of Yang-Mills with a finite gauge group, 
with fields dependent only upon time, ({\ref{doulamor}), 
characteristic features of chaotic behaviour have been demonstrated for the
solutions of the second-order equations of motion \cite{savvidy}.
It would be interesting to also know what is the situation for the behaviour of
solutions to the second-order dynamical equations of the PB 
system (\ref{legba}).

\end{document}